\documentclass[showpacs,preprintnumbers,twocolumn,superscriptaddress,aps,nofootinbib]{revtex4}
\usepackage{amssymb}
\usepackage[centertags]{amsmath}
\usepackage{txfonts}
\usepackage{epsfig}
\usepackage{bm}
\usepackage{color}
\usepackage{graphicx,graphics}
\usepackage{multirow}
\usepackage{float}
\usepackage{ulem}
\usepackage{appendix}
\usepackage[pdfstartview=FitH]{hyperref}
\allowdisplaybreaks[2]

\begin{document}

\title{Charmed hadron production via equal-velocity quark combination in ultra-relativistic heavy ion collisions}

\author{Rui-Qin Wang}
\email {wangruiqin2010@mail.sdu.edu.cn}
\affiliation{School of Physics and Physical Engineering, Qufu Normal University, Shandong 273165, China}

\author{Jun Song}
\email {songjun2011@jnxy.edu.cn}
\affiliation{Department of Physics, Jining University, Shandong 273155, China}

\author{Feng-Lan Shao}
\email {shaofl@mail.sdu.edu.cn}
\affiliation{School of Physics and Physical Engineering, Qufu Normal University, Shandong 273165, China}

\author{Zuo-Tang Liang}
\email {liang@sdu.edu.cn}
\affiliation{Key Laboratory of Particle Physics and Particle Irradiation (MOE), 
Institute of Frontier and Interdisciplinary Science, Shandong University, Qingdao, Shandong 266237, China}

\begin{abstract}

Recent data on the production of $D$ mesons and $\Lambda_c^+$ baryons 
in heavy ion collisions at the Relativistic Heavy Ion Collider 
and the Large Hadron Collider exhibit a number of striking characteristics 
such as enhanced yield ratios $D_s^+/D^0$, $\Lambda_c^+/D^0$ 
and their transverse momentum dependences.
In this paper, we derive the momentum dependence of open charm mesons 
and singly charmed baryons produced in ultra-relativistic heavy ion collisions via the equal-velocity quark combination.
We present analytic expressions and numerical results of yield ratios and compare them with the experimental data available. 
We make predictions for other charmed hadrons.

\end{abstract}

\maketitle

\section{introduction}

In ultra-relativistic heavy ion collisions, heavy quarks and anti-quarks are produced predominantly via initial hard scatterings and then experience the entire evolution of a violently interacting medium of deconfined quarks and gluons --- the quark gluon plasma (QGP). 
These heavy flavor quarks and anti-quarks interact strongly with constituents of the QGP medium~\cite{2016EPJC,2016JPG}, 
exchange energy and momentum in the partonic evolution process~\cite{exchangeEP1,exchangeEP2,ClossE2001PLB} and combine with them to form heavy flavor hadrons that can be observed experimentally. 
Therefore, the production of heavy flavor hadrons is usually believed to play a very special role in studying the hadronization mechanism and properties of the QGP matter~\cite{2016PRep,Batsouli2003PLB,NuXuNPA2003,ZWLinPRC2003,MoorePRC2005,HeesPRC2006}.

Recently, experimental data on open charm mesons and $\Lambda_c^+$ baryons with improved precision 
and extended transverse momentum ($p_T$) coverage at the Relativistic Heavy Ion Collider (RHIC) 
and the Large Hadron Collider (LHC) have been 
published~\cite{D0PRC2019,502PbPb2018JHEP,LcD0NPA2019,DsD0NPA2017,DsDLHCarXiv2019,LcD2019PLB,pppPb2018JHEP}. 
These data show indeed a number of fascinating features. 
The most striking ones might be the enhancement of the strange to non-strange meson yield ratio $D_s^+/D$ and that of the baryon-to-meson ratio $\Lambda_c^+/D^0$. 
It has been observed~\cite{DsD0NPA2017,DsDLHCarXiv2019} that the $D_s^+/D$ ratio in ultra-relativistic heavy ion collisions 
is much higher than predictions of fragmentation models~\cite{fragDsD2016EPJC,fragDsD2006JHEP}.
The $\Lambda_c^+/D^0$ ratio exhibits also a very strong enhancement trend at intermediate $p_T$ 
area not only in RHIC Au-Au and LHC Pb-Pb collisions but also in $pp$ and $p$-Pb reactions at the extremely high LHC energies~\cite{pppPb2018JHEP}.
They are much higher than the predictions of perturbative QCD
and event generators based on fragmentation models~\cite{pppPb2018JHEP,Fragm2018PRD}.

Theoretically, much effort has also been made to the hadronization mechanism of charm quarks in high energy 
heavy ion collisions~\cite{statis2008PLB,statis2007EPJC,Creview2016EPJC,DAAPLB2004,Lamc-diquark,Bccombi2013PRC,charmCombFrag,HJXu2013,RLcD2018EPJC,PFZhuang2019arXiv,MHe2019arXiv,KJSun2019arXiv,RQWang2015PRC}.
It seems in particular that the coalescence or (re-)combination mechanism should play an irreplaceable role 
in describing the production of charmed hadrons with low and intermediate transverse momenta~\cite{Creview2016EPJC,DAAPLB2004,Lamc-diquark,Bccombi2013PRC,charmCombFrag,HJXu2013,RLcD2018EPJC,PFZhuang2019arXiv,MHe2019arXiv,KJSun2019arXiv}.
In this connection, we note in particular that  
the quark combination model under equal-velocity combination (hereafter referred as EVC)~\cite{equal-velocity} 
provides a very simple and elegant way to describe the $p_T$ dependence of production rates of hadrons and 
has been successfully applied~\cite{HHLi2018PRC,JSong2018EPJC}  to $pp$ and $p$-Pb 
collisions to describe the enhancement of $\Lambda_c^+/D^0$. 
It is therefore interesting to see whether it can also be applied to heavy ion collisions to describe the heavy flavor hadron production as well.  

In this paper, we apply the quark combination via EVC to ultra-relativistic heavy ion collisions to study the
production of open charm $D$ mesons and singly charmed baryons in the low and intermediate $p_T$ regions. 
We present the detailed derivations and give analytic expressions for $p_T$-dependences of ratios of production rates 
of different hadrons. 
We compare the results with the experimental data available~\cite{DsD0NPA2017,LcD0NPA2019,D0PRC2019} and make predictions for other types of hadrons.  

The rest of the paper is organized as follows. 
In Sec.~II, we present the derivation of the momentum-dependent production of singly charmed hadrons 
in the quark combination model via EVC in ultra-relativistic heavy ion collisions. 
We present in particular the analytic expressions for two kinds of yield ratios measured by RHIC and LHC experiments~\cite{DsD0NPA2017,DsDLHCarXiv2019,LcD0NPA2019,LcD2019PLB} 
as the function of $p_T$ and discuss their qualitative properties. 
In Sec.~III, we apply the results obtained in Sec.~II to Au-Au collisions at $\sqrt{s_{NN}}= 200$ GeV.
In Sec.~IV, we present a summary.

\section{The charmed hadron production in the quark combination via EVC}   \label{model}

The basic idea of the quark combination mechanism and formulae for the momentum dependence have been presented in many different literatures. 
An example for the derivation for light flavor hadrons starting from the general formulae has been given in~\cite{RQWang2012PRC}  
where $p_T$-integrated yield correlations have been calculated. 
Here, in this section, for explicitness, we just follow the same procedure as that in~\cite{RQWang2012PRC} 
and present results of momentum dependence for singly charmed hadrons. 
After that, we present the corresponding results obtained under EVC. 

\subsection{The general formalism}  

To derive the momentum-dependent production rates of mesons and baryons in ultra-relativistic heavy ion collisions, 
we just as in~\cite{RQWang2012PRC} start with a color-neutral quark-anti-quark system with $N_{q_i}$ quarks 
of flavor $q_i$ and $N_{\bar q_i}$ anti-quarks of flavor $\bar q_i$ and suppose they hadronize via the quark combination mechanism.
The momentum distributions $f_{M_j}(p;N_{q_i},N_{\bar{q}_i})$ and $f_{B_j}(p;N_{q_i},N_{\bar{q}_i})$ 
for the directly produced meson $M_j$  
and baryon $B_j$ are given by,
{\setlength\arraycolsep{0.2pt}
\begin{eqnarray}
f_{M_j}(p;N_{q_i},N_{\bar{q}_i})&&= \sum\limits_{\bar{q}_{1}q_{2}}  \int dp_1 dp_2 N_{\bar{q}_{1}q_{2}} f^{(n)}_{\bar{q}_{1}q_{2}}(p_1,p_2;N_{q_i},N_{\bar{q}_i})  \nonumber  \\
&&~~~~~~\times \mathcal {R}_{M_j,\bar{q}_{1}q_{2}}(p,p_1,p_2;N_{q_i},N_{\bar{q}_i}),    \label{eq:fDjgeneral}     \\
% ---------------------------------------------------------------------------------------------------
f_{B_j}(p;N_{q_i},N_{\bar{q}_i})&&= \sum\limits_{q_{1}q_{2}q_{3}}  \int dp_1 dp_2 dp_3   \nonumber  \\
&&~~~~~~ \times N_{q_1q_2q_3} f^{(n)}_{q_{1}q_{2}q_{3}}(p_1,p_2,p_3;N_{q_i},N_{\bar{q}_i})   \nonumber  \\
&&~~~~~~ \times \mathcal {R}_{B_j,q_{1}q_{2}q_{3}}(p,p_1,p_2,p_3;N_{q_i},N_{\bar{q}_i}),    \label{eq:fBcjgeneral}
\end{eqnarray} }%
where $f^{(n)}_{\bar{q}_{1}q_{2}}$ and $f^{(n)}_{q_{1}q_{2}q_{3}}$ are normalized joint
momentum distributions for $\bar{q}_{1}q_{2}$ and $q_{1}q_{2}q_{3}$ respectively;
$N_{\bar{q}_{1}q_{2}}= N_{\bar q_1 }N_{q _2}$ is the number of $\bar q_1 q_2$ pairs; 
$N_{q_1 q_2 q_3}$ is the number of $q_1 q_2 q_3$ clusters in the system, and 
it takes $N_{q_1}N_{q_2}N_{q_3}$ for $q_1\neq q_2 \neq q_3$, 
$N_{q_1}(N_{q_1}-1)N_{q_3}$ for $q_1 = q_2 \neq q_3$
and $N_{q_1}(N_{q_1}-1)(N_{q_1}-2)$ for $q_1= q_2 = q_3$; 
kernel functions $\mathcal {R}_{M_j,\bar{q}_{1}q_{2}}$ and $\mathcal {R}_{B_j,q_{1}q_{2}q_{3}}$ 
stand for the probability density for a $\bar q_1q_2$ pair with momenta $p_1$ and $p_2$ 
to combine into a meson $M_j$ of momentum $p$
and that for a $q_1q_2q_3$ cluster with $p_1$, $p_2$ and $p_3$ to combine into a baryon $B_j$ of $p$ respectively. 

Just as discussed in~\cite{RQWang2012PRC},  
$\mathcal {R}_{M_j,\bar{q}_{1}q_{2}}$ and $\mathcal {R}_{B_j,q_{1}q_{2}q_{3}}$ 
carry the kinematical and dynamical information of the quark combination.
Their precise forms can not be derived from the first principles due to their complicated non-perturbative nature.
Nevertheless, they are constrained by a number of symmetry laws and rules 
such as the momentum conservation, 
constraints due to intrinsic quantum numbers such as spin and flavor,  
the requirement of the hadronization unitarity so that the production of all the open and hidden charm hadrons 
should exhaust all charm quarks and anti-quarks in the system, 
and the meson-baryon production competition and so on. 
To take these constraints into account explicitly, 
we re-write them in the following forms, 
{\setlength\arraycolsep{0.2pt}
\begin{eqnarray}
&& \mathcal {R}_{M_j,\bar{q}_{1}q_{2}}(p,p_1,p_2;N_{q_i},N_{\bar{q}_i}) = C_{M_j}  \mathcal {R}_{\bar q_{1} q_2}^{(f)}   \nonumber  \\
&&~~~\times \mathcal {R}_{M,\bar{q}_{1}q_{2}}(p_1,p_2;N_{q_i},N_{\bar{q}_i})   \delta(p_1+p_2-p), \label{eq:RDjgeneral} \\
%----------------------------------------------------------------------------------------------------------------------------------------------
&& \mathcal {R}_{B_j,q_1q_2q_3}(p,p_1,p_2,p_3;N_{q_i},N_{\bar{q}_i})= C_{B_j}   \mathcal {R}_{q_{1}q_{2}q_3}^{(f)}   \nonumber  \\
&&~~~\times \mathcal {R}_{B,q_{1}q_{2}q_{3}}(p_1,p_2,p_3;N_{q_i},N_{\bar{q}_i}) \delta(p_1+p_2+p_3-p). ~~\label{eq:RBcjgeneral} 
\end{eqnarray} }%
Here the $\delta$-functions are used to guarantee the momentum conservation. 
The factors $\mathcal {R}_{\bar q_1q_2}^{(f)}$ and $\mathcal {R}_{q_{1}q_{2}q_3}^{(f)}$ contain Kronecker $\delta$'s 
to guarantee the quark flavor conservation, e.g., if $M_j$ is a $D$-meson with constituent quark content $\bar q c$,  
$\mathcal {R}^{(f)}_{\bar q_{1} q_2}=\delta_{q_1,q} \delta_{q_2,c}$. 
For $\mathcal {R}_{q_{1}q_{2}q_3}^{(f)}$, we need to include a number $N_{iter}$ to account for the fact that 
there are different iterations for the flavors of the three quarks 
and $N_{iter}=1$, $3$ or $6$ for $q_1=q_2=q_3$, $q_1=q_2\not=q_3$ or $q_1\not=q_2\not=q_3$.

The factor $C_{M_j}$ is the probability for $M$ to be $M_j$ 
if the quark content of $M$ is the same as $M_j$ and similar for $C_{B_j}$.
In the case that only $J^P=0^-$ and $1^-$ mesons and $J^P=(1/2)^+$ and $(3/2)^+$ baryons are considered, 
they are just determined completely by 
the production ratio of vector to pseudo-scalar mesons
and that of $J^P=({1}/{2})^+$ to $J^P=({3}/{2})^+$ baryons with the same flavor content. 

The remaining factor $\mathcal {R}_{M,\bar{q}_{1}q_{2}}$ 
now stands for the probability of a $\bar q_1q_2$ pair with momenta $p_1$ and $p_2$ to combine into a meson $M$ 
with any momentum and other quantum numbers and similar for $\mathcal {R}_{B,q_{1}q_{2}q_{3}}$.
They depend on the momenta of the (anti-)quarks and their situated environments represented by $N_{q_i}$ and $N_{\bar{q}_i}$ 
and should be determined by the dynamics in the combination process.

In this way, we obtain the momentum distribution for a charmed meson $D_j$ with quark flavor content $\bar q_1c$ 
and that for a singly charmed baryon $B_j^c$ with $q_1q_2c$ as, 
{\setlength\arraycolsep{0.2pt}
\begin{eqnarray}
&& f_{D_j}(p;N_{q_i},N_{\bar{q}_i})= \int dp_1 dp_2
      N_{\bar{q}_{1}c} f^{(n)}_{\bar{q}_{1}c}(p_1,p_2;N_{q_i},N_{\bar{q}_i})  \nonumber  \\
 &&~~~~~~\times
      C_{D_j}  \mathcal {R}_{M,\bar{q}_1c}(p_1,p_2;N_{q_i},N_{\bar{q}_i})  \delta(p_1+p_2-p),   \label{eq:fDj0}     \\
% ---------------------------------------------------------------------------------------------------
&& f_{B_j^c}(p;N_{q_i},N_{\bar{q}_i})=  \int dp_1 dp_2 dp_3  \nonumber\\ 
&&~~~~~~\times N_{q_1q_2c} f^{(n)}_{q_{1}q_{2}c}(p_1,p_2,p_3;N_{q_i},N_{\bar{q}_i})  C_{B^c_j}   N_{iter} \nonumber\\
&&~~~~~~\times   \mathcal {R}_{B,q_{1}q_{2}c}(p_1,p_2,p_3;N_{q_i},N_{\bar{q}_i}) \delta(p_1+p_2+p_3-p).~~~~~      \label{eq:fBcj0}
\end{eqnarray} }%
Here, as well as in the following of this paper, when talking about charmed mesons $D_j$'s and singly charmed baryons $B_j^c$'s, 
$q_1$, $q_2$ and $\bar q_1$ denote $u$, $d$ and $s$ flavors of quarks or anti-quarks.   

\subsection{The momentum distribution under EVC}

It has been shown that~\cite{equal-velocity} EVC seems to work well for light and strange hadron production in low and intermediate $p_T$ regions. 
For charmed hadrons, the mass differences between the charm and other quarks are large. 
It is therefore much more sensitive and more interesting to see whether the combination proceeds via EVC or other rules.  

Under EVC, $\mathcal {R}_{M,\bar q_1q_2}$ and $\mathcal {R}_{B,q_1q_2q_3}$ take, 
{\setlength\arraycolsep{0.2pt}
\begin{eqnarray}
&& \mathcal {R}_{M,\bar q_1q_2} (p_{1},p_{2};N_{q_i},N_{\bar q_i})
= \mathcal A_{M,{\bar q_1q_2}} (N_{q_i},N_{\bar q_i}) \nonumber  \\
                       &&~~~~~~~\times \frac{m_{q_1}+m_{q_2}}{m_{q_1}m_{q_2} }
\delta(\frac{p_{1}}{m_{q_1}}-\frac{p_{2}}{m_{q_2}}),   
                       \label{eq:equalvRM}    \\
&& \mathcal {R}_{B,q_1q_2q_3} (p_{1},p_{2},p_{3};N_{q_i},N_{\bar q_i})
= \mathcal A_{B,{q_1q_2q_3}} (N_{q_i},N_{\bar q_i})  \nonumber  \\
                       &&~~~~~~~\times \frac{m_{q_1}+m_{q_2}+m_{q_3}}{m_{q_1} m_{q_2} m_{q_3}}
    \delta(\frac{p_{1}}{m_{q_1}}-\frac{p_{2}}{m_{q_2}}) \delta(\frac{p_{1}}{m_{q_1}}-\frac{p_{3}}{m_{q_3}}), ~~~ \label{eq:equalvRB}
\end{eqnarray} }%
where $m_{q_i}$ is the constituent quark mass and is taken as $m_u=m_d=0.33$ GeV, $m_s=0.5$ GeV and $m_c=1.5$ GeV 
in the numerical calculations in this paper; 
$\mathcal A_{M,{\bar q_1q_2}}$ and $\mathcal A_{B,{q_1q_2q_3}}$ 
are factors denoting the meson-baryon production competition as well as guaranteeing the unitarity.
They depend on the numbers of different flavor quarks and anti-quarks in the considered system.
The mass term is introduced for normalization, so that the original kernel functions 
in Eqs.~(\ref{eq:fDjgeneral}) and (\ref{eq:fBcjgeneral}) take the form,
{\setlength\arraycolsep{0.2pt}
\begin{eqnarray}
&& \mathcal {R}_{M_j,\bar q_1q_2} (p_{1},p_{2};N_{q_i},N_{\bar q_i})
= C_{M_j}  \mathcal {R}_{\bar q_{1} q_2}^{(f)} \mathcal A_{M,{\bar q_1q_2}} (N_{q_i},N_{\bar q_i}) \nonumber  \\
                       &&~~~~~~~~~~~~~~~~~~~~~\times \delta(p_1-x_{q_1q_2}^{q_1}p) \delta(p_2-x_{q_1q_2}^{q_2}p),   \\
%---------------------------------------------------------------------------------------------------------------                       
&& \mathcal {R}_{B_j,q_1q_2q_3} (p_{1},p_{2},p_{3};N_{q_i},N_{\bar q_i})
= C_{B_j}   \mathcal {R}_{q_{1}q_{2}q_3}^{(f)} \nonumber  \\
 &&~~~~~~~~~~~~~~~~~~~~~\times   \mathcal A_{B,{q_1q_2q_3}} (N_{q_i},N_{\bar q_i})  \delta(p_1-x_{q_1q_2q_3}^{q_1}p)   \nonumber  \\
                       &&~~~~~~~~~~~~~~~~~~~~~\times \delta(p_2-x_{q_1q_2q_3}^{q_2}p)
                       \delta(p_3-x_{q_1q_2q_3}^{q_3}p), ~~~ 
\end{eqnarray} }%
where $x^{q_i}_{q_1q_2}=m_{q_i}/(m_{q_1}+m_{q_2})$ and 
$x^{q_i}_{q_1q_2q_3}=m_{q_i}/(m_{q_1}+m_{q_2}+m_{q_3})$ are the fraction of the momentum of 
the produced hadron carried by $q_i$. 

Substituting Eqs.~(\ref{eq:equalvRM}) and (\ref{eq:equalvRB}) into Eqs.~(\ref{eq:fDj0}) and (\ref{eq:fBcj0}) 
and carrying out the integration over the quark momenta, we obtain,
{\setlength\arraycolsep{0.2pt}
\begin{eqnarray}
f_{D_j}(p) =&&  C_{D_j} \mathcal A_{M,{\bar q_1c}} (N_{q_i},N_{\bar q_i}) 
                N_{\bar q_1c} f^{(n)}_{\bar q_1 c} (x_{q_1c}^{q_1} p,x_{q_1c}^{c}p), ~~~                       
\label{eq:fDj} \\
% -------------------------------------------------------------------------------------
 f_{B^c_j}(p) =&&C_{B^c_j}  \mathcal A_{B,{q_1q_2c}} (N_{q_i},N_{\bar q_i}) N_{q_1q_2c} N_{iter} \nonumber\\ 
 && \times  f^{(n)}_{q_{1}q_{2}c}(x_{q_1q_2c}^{q_1} p,x_{q_1q_2c}^{q_2}p,x_{q_1q_2c}^{c}p).    \label{eq:fBcj}
\end{eqnarray} }%
Here and from now on in this paper, 
we suppress the arguments $N_{q_i}, N_{\bar{q}_i}$ in the momentum distribution functions for explicitness.

The factor $\mathcal A_{M,{\bar q_1c}} (N_{q_i},N_{\bar q_i})$ 
is the probability for a charm quark to capture a specific anti-quark $\bar q_1$ 
to form a meson in the bulk quark-anti-quark system, 
it should be inversely proportional to the total number of quarks and anti-quarks $N_q+N_{\bar q}$.
Similarly, $\mathcal A_{B,{q_{1}q_{2}c}} (N_{q_i},N_{\bar q_i})$ 
should be proportional to ${1}/ {(N_q+N_{\bar q})^2}$.
Therefore, one can write 
{\setlength\arraycolsep{0.2pt}
\begin{eqnarray}
&& \mathcal A_{M,{\bar q_1}c} (N_{q_i},N_{\bar q_i}) = \mathcal A_{M}/ (N_q+N_{\bar q}),  \label{eq:AM}  \\
&& \mathcal A_{B,q_{1}q_{2}c} (N_{q_i},N_{\bar q_i}) = \mathcal A_{B}/ (N_q+N_{\bar q})^2, \label{eq:AB}
\end{eqnarray} }%
where $\mathcal A_{M}$ and $\mathcal A_{B}$ are proportionality coefficients 
and they closely relate to the unitarity and the meson-baryon production competition. 
For a given quark-anti-quark system, $\mathcal A_{M}$ and $\mathcal A_{B}$ should be universal for all different $D$ mesons 
and singly charmed baryons due to the quark flavor blindness of the strong interaction.
Substituting Eqs.~(\ref{eq:AM}) and (\ref{eq:AB}) into Eqs.~(\ref{eq:fDj}) and (\ref{eq:fBcj}), respectively, we have
{\setlength\arraycolsep{0.2pt}
\begin{eqnarray}
 f_{D_j}(p) =&& N_c \mathcal A_{M} C_{D_j}  \lambda_{\bar q_1} 
f^{(n)}_{\bar q_1c} (x_{q_1c}^{q_1} p,x_{q_1c}^{c} p),  \label{eq:fDj2}     \\
% ---------------------------------------------------------------------------------------------------
 f_{B^c_j}(p) =&& N_c  \mathcal A_{B} C_{B^c_j}  \lambda_{q_1q_2} N_{iter} \nonumber\\
  && \times 
  f^{(n)}_{q_1q_2c}(x_{q_1q_2c}^{q_1}p,x_{q_1q_2c}^{q_2}p,x_{q_1q_2c}^{c}p), ~~~\label{eq:fBcj2}
\end{eqnarray} }%
where $\lambda_{\bar q_1}\equiv N_{\bar q_1}/({N_q+N_{\bar q}})$ and  
$\lambda_{q_1q_2}\equiv N_{q_1q_2}/{(N_q+N_{\bar q})^2}$. 
If we consider a quark-anti-quark system in the mid-rapidity region at high energies 
so that the influence of net quarks from the colliding nuclei can be neglected,
the ratio $\lambda_{\bar q_1}$ and $\lambda_{q_1q_2}$ are 
both determined completely by the strangeness suppression factor $\lambda_s$. 

If we neglect correlations of momentum distributions of quarks and/or anti-quarks of different flavors 
in the system, i.e. we take, 
{\setlength\arraycolsep{0.2pt}
\begin{eqnarray}
&& f^{(n)}_{\bar{q}_{1}q_{2}}(p_1,p_2) 
= f^{(n)}_{\bar{q}_{1}}(p_1) f^{(n)}_{q_{2}}(p_2),   \label{eq:fnq1q2}   \\ 
&& f^{(n)}_{q_{1}q_{2}q_{3}}(p_1,p_2,p_3) 
=f^{(n)}_{q_{1}}(p_1)  f^{(n)}_{q_{2}}(p_2)  f^{(n)}_{q_{3}}(p_3). ~~~~~~~  \label{eq:fnq1q2q3}
\end{eqnarray} }%
In this case, we obtain 
{\setlength\arraycolsep{0.2pt}
\begin{eqnarray}
 f_{D_j}(p) =&& N_c \mathcal A_{M} C_{D_j}  \lambda_{\bar{q}_{1}}  
 f^{(n)}_{\bar{q}_{1}}(x_{q_1c}^{q_1}p)   f^{(n)}_{c}(x_{q_1c}^{c}p),~~~~~~     \label{eq:fDj3}     \\
% ---------------------------------------------------------------------------------------------------
 f_{B^c_j}(p) =&& N_c \mathcal A_{B} C_{B^c_j}  \lambda_{q_{1}q_{2}}  N_{iter} \nonumber\\
 && \times f^{(n)}_{q_{1}}(x_{q_1q_2c}^{q_1}p) f^{(n)}_{q_{2}}(x_{q_1q_2c}^{q_2}p)  f^{(n)}_{c}(x_{q_1q_2c}^{c}p).     \label{eq:fBcj3}
\end{eqnarray} }%

By using Eqs.~(\ref{eq:fDj3}) and (\ref{eq:fBcj3}), we can 
calculate momentum distributions and ratios for the production of different charmed hadrons. 
The factors $\mathcal A_{M}$ and $\mathcal A_{B}$ 
are determined by charm quark number conservation in the combination process 
and the charmed baryon-to-meson production ratio $N_{B^c}/N_D$. 
Here $N_D$ and $N_{B^c}$ are the total number of all the produced $D$ mesons and that of singly charmed baryons. 
For the charm quark number conservation, we need in principle to consider 
the production of hadrons besides singly charmed ones such as charmonia, doubly and triply charmed baryons and even exotic states. 
However, the production rates for them are very small and they exhaust about less than 5\% of total charm quarks~\cite{XDong2019arXiv}.
In the numerical calculations in the following of this paper, we will just neglect them 
so the charm quark number conservation just takes $N_D+N_{B^c}\approx N_c$.   
In this approximation, $\mathcal A_{M}$ and $\mathcal A_{B}$  are just determined by the ratio $R^{(c)}_{B/M}\equiv N_{B^c}/N_D$ 
that is taken as a parameter fixed by the experimental data of one yield ratio such as $\Lambda_c^+/D^0$ in the calculations. 

\subsection{Decay contributions}

To compare with the experimental data, 
we need to include strong and electromagnetic decay contributions from short-lived charmed resonances~\cite{PDG2018}. 
In our case, we need only to consider decays of $D^*$ mesons, $J^P=(3/2)^+$ singly charmed baryons and $\Sigma_c$ baryons. 
They all decay into a $D$ meson or a $J^P=(1/2)^+$ singly charmed baryon with a light particle such as a pion or a photon.  
In such a decay process, the momentum of the light daughter particle is so small that can be neglected compared 
to that of the heavy daughter charmed hadron. 
We can approximately take the momentum of the daughter charmed hadron equal to that of the mother charmed hadron. 
In this approximation, and take vector to pseudo-scalar meson production ratio as 1.5~\cite{RVP2012JHEP,HHLi2018PRC}, 
we find the results for the final $D$ mesons as,
{\setlength\arraycolsep{0.2pt}
\begin{eqnarray}
 && f_{D^0}^{(fin)}(p) \approx 3.516 f_{D^0}(p),  \label{eq:D0fin}   \\
 && f_{D^+}^{(fin)}(p)  \approx 1.485 f_{D^+}(p),  \label{eq:Dpfin}  \\
 && f_{D_s^+}^{(fin)}(p)  \approx 2.5 f_{D_s^+}(p).  \label{eq:Dspfin} 
\end{eqnarray}  }%
Similarly, we take $J^P=(1/2)^+$ to $J^P=(3/2)^+$ singly charmed baryon ratio as 2~\cite{HHLi2018PRC}, 
and obtain,  
{\setlength\arraycolsep{0.2pt}
\begin{eqnarray}
 && f_{\Lambda_c^+}^{(fin)}(p)  \approx 5 f_{\Lambda_c^+}(p),  \label{eq:Lamcpfin} \\
 && f_{\Sigma_c^0}^{(fin)}(p)  \approx  f_{\Sigma_c^0}(p),  \label{eq:Sigc0fin} \\
 && f_{\Sigma_c^+}^{(fin)}(p)  \approx f_{\Sigma_c^+}(p),  \label{eq:Sigcpfin} \\
 && f_{\Sigma_c^{++}}^{(fin)}(p)  \approx f_{\Sigma_c^{++}}(p),  \label{eq:Sigcppfin} \\
 && f_{\Xi_c^0}^{(fin)}(p)  \approx 2.5 f_{\Xi_c^0}(p),  \label{eq:Xic0fin} \\
 && f_{\Xi_c^+}^{(fin)}(p)  \approx 2.5 f_{\Xi_c^+}(p),  \label{eq:Xicpfin} \\
 && f_{\Omega_c^0}^{(fin)}(p)  \approx 1.5 f_{\Omega_c^0}(p).  \label{eq:Omec0fin}
\end{eqnarray}  }%
From these results, we see clearly that contributions from resonance decays are important 
for most of the charmed hadrons.
For example, for $\Lambda_c^+$ baryons, 
about 80\% are from decay contributions and only about 20\% are directly produced ones.

\subsection{Ratios of different hadrons}

We consider the production of charmed hadrons at midrapidity $y=0$
and apply Eqs.~(\ref{eq:fDj3}) and (\ref{eq:fBcj3}) to obtain the $p_T$ dependence. 
From them, we calculate ratios of different charmed hadrons.
In this case, the numbers of different flavor quarks are just replaced by the number densities $dN_{q_i}/dy$ at $y=0$.

We first consider yield ratios of strange to non-strange hadrons in the charm sector. 
They are given by,
{\setlength\arraycolsep{0.2pt}
\begin{eqnarray}
 &&\frac{D_s^+}{D^0}  = 0.711\lambda_s \frac{f^{(n)}_{\bar s}(x^{s}_{sc}p_T) f^{(n)}_{c}(x^{c}_{sc}p_T)}
                                        {f^{(n)}_{\bar d}(x^{d}_{dc}p_T) f^{(n)}_{c}(x^{c}_{dc}p_T)},  \label{eq:RDsD0fin}   \\
%----------------------------------------------------------------------------------------                                           
&&\frac{\Xi_c^+}{\Lambda_c^+} 
  = 0.5\lambda_s \frac{f^{(n)}_{d}(x^{d}_{dsc}p_T) f^{(n)}_{s}(x^{s}_{dsc}p_T) f^{(n)}_{c}(x^{c}_{dsc}p_T)}
                        {[f^{(n)}_{d}(x^{d}_{ddc}p_T)]^2 f^{(n)}_{c}(x^{c}_{ddc}p_T)},  \label{eq:RXicLamcfin}   \\                                           
%---------------------------------------------------------------------------------------- 
&&\frac{\Omega_c^0}{\Xi_c^0}  
  = 0.5\lambda_s \frac{[f^{(n)}_{s}(x^{s}_{ssc}p_T)]^2 f^{(n)}_{c}(x^{c}_{ssc}p_T)}
                      {f^{(n)}_{d}(x^{d}_{dsc}p_T) f^{(n)}_{s}(x^{s}_{dsc}p_T) f^{(n)}_{c}(x^{c}_{dsc}p_T)}, ~~~~~~~~~ \label{eq:ROmecXicfin}   \\                                           
%---------------------------------------------------------------------------------------- 
&&\frac{\Omega_c^0}{\Lambda_c^+} 
   = 0.25\lambda_s^2 \frac{[f^{(n)}_{s}(x^{s}_{ssc}p_T)]^2 f^{(n)}_{c}(x^{c}_{ssc}p_T)}
                      {[f^{(n)}_{d}(x^{d}_{ddc}p_T)]^2  f^{(n)}_{c}(x^{c}_{ddc}p_T)}.  \label{eq:ROmecLamcfin}
\end{eqnarray}  }%
We can use them to calculate these yield ratios numerically. 
Here we see clearly that these ratios depend not only on the strangeness suppression factor $\lambda_s$ 
but also on the ratios of $p_T$-distributions of different flavors of (anti-)quarks. 

To see the qualitative features more explicitly, we note that, 
since $m_c$ is much larger than $m_d$ and $m_s$, it dominates 
the sum of $m_c$ with other quark mass. 
We have that 
$x^{q_1}_{q_1q_1c}\sim x^{q_1}_{q_1q_2c}\sim x^{q_1}_{q_1c}$,  
$x^{c}_{q_1q_2c}\sim x^{c}_{q_1c}\sim x^{c}_{q_2c}$ 
and the latter should be much larger than the former.
We take, in a rough approximation, the former as the same,  
and re-write Eqs.~(\ref{eq:RDsD0fin}-\ref{eq:ROmecLamcfin}) in the form, 
{\setlength\arraycolsep{0.2pt}
\begin{eqnarray}
 &&\frac{D_s^+}{D^0} = 0.711\lambda_s \frac{f^{(n)}_{\bar s}(x^{s}_{sc}p_T)}
                                        {f^{(n)}_{\bar d}(x^{d}_{dc}p_T)}  \frac{f^{(n)}_{c}(x^{c}_{sc}p_T)}
                                        {f^{(n)}_{c}(x^{c}_{dc}p_T)},  \label{eq:RDsD0rough}   \\
%----------------------------------------------------------------------------------------                                           
&&\frac{\Xi_c^+}{\Lambda_c^+} 
  \approx 0.5\lambda_s \frac{f^{(n)}_{s}(x^{s}_{dsc}p_T) }
                        {f^{(n)}_{d}(x^{d}_{ddc}p_T) } \frac{f^{(n)}_{c}(x^{c}_{dsc}p_T)}
                        {f^{(n)}_{c}(x^{c}_{ddc}p_T)},  \label{eq:RXicLamcrough}   \\                                           
%---------------------------------------------------------------------------------------- 
&&\frac{\Omega_c^0}{\Xi_c^0} 
  \approx 0.5\lambda_s \frac{f^{(n)}_{s}(x^{s}_{ssc}p_T) }
                      {f^{(n)}_{d}(x^{d}_{dsc}p_T)  }
                      \frac{f^{(n)}_{c}(x^{c}_{ssc}p_T)}
                      {f^{(n)}_{c}(x^{c}_{dsc}p_T)}, \label{eq:ROmecXicrough}   \\                                           
%---------------------------------------------------------------------------------------- 
&&\frac{\Omega_c^0}{\Lambda_c^+} 
   = 0.25\lambda_s^2 \left[\frac{f^{(n)}_{s}(x^{s}_{ssc}p_T)}
                      {f^{(n)}_{d}(x^{d}_{ddc}p_T)}\right]^2 
                      \frac{f^{(n)}_{c}(x^{c}_{ssc}p_T)}
                      {f^{(n)}_{c}(x^{c}_{ddc}p_T)}.  ~~~~~~ \label{eq:ROmecLamcrough}
\end{eqnarray}  }%
Here, we see clearly that, besides $\lambda_s$ (or $\lambda_s^2$), 
these ratios are proportional to the ratio of $s$ to $d$-quark spectrum (or squared) 
and the ratio of $c$-quark spectrum at slightly different $p_T$ values. 
Since the values of $x$'s involved here are all quite small for $d$ and $s$ quarks, they should be sensitive to 
the $d$- and $s$-quark $p_T$ spectra in the relatively low $p_T$ regions. 
The observed enhancement of strange to non-strange charmed hadron ratios 
does not necessarily come from the enhancement of $\lambda_s$ 
but can also from the influence of the $p_T$-spectra of quarks. 
 
Similarly, for baryon-to-meson ratios,  
we obtain from Eqs.~(\ref{eq:D0fin}-\ref{eq:Omec0fin}) that, 
{\setlength\arraycolsep{0.2pt}
\begin{eqnarray}
&& \frac{\Lambda_c^+}{D^0} = \frac{4.267}{2+\lambda_s} \frac{\mathcal A_{B}}{\mathcal A_{M}}
                \frac{ [f^{(n)}_{d}(x^{d}_{ddc}p_T)]^2  f^{(n)}_{c}(x^{c}_{ddc}p_T)}
                     {f^{(n)}_{\bar d}(x^{d}_{dc}p_T) f^{(n)}_{c}(x^{c}_{dc}p_T)},               \label{eq:RLcpD0fin}   \\
%-------------------------------------------------------------------------------------------------------------------------------                     
&& \frac{\Sigma_c^0}{D^0} =  \frac{0.711}{2+\lambda_s} \frac{\mathcal A_{B}}{\mathcal A_{M}}
                \frac{ [f^{(n)}_{d}(x^{d}_{ddc}p_T)]^2 f^{(n)}_{c}(x^{c}_{ddc}p_T)}
                      {f^{(n)}_{\bar d}(x^{d}_{dc}p_T) f^{(n)}_{c}(x^{c}_{dc}p_T)},           \label{eq:RSc0D0fin}     \\  
%-------------------------------------------------------------------------------------------------------------------------------                      
&& \frac{\Sigma_c^{++}}{D^+}  =  \frac{1.684}{2+\lambda_s} \frac{\mathcal A_{B}}{\mathcal A_{M}}
                    \frac{ [f^{(n)}_{d}(x^{d}_{ddc}p_T)]^2 f^{(n)}_{c}(x^{c}_{ddc}p_T)}
                         {f^{(n)}_{\bar d}(x^{d}_{dc}p_T) f^{(n)}_{c}(x^{c}_{dc}p_T)},            \label{eq:RScppDpfin}   \\                                         
%-------------------------------------------------------------------------------------------------------------------------------
&& \frac{\Xi_c^0}{D^0}  =  \frac{2.133\lambda_s}{2+\lambda_s} \frac{\mathcal A_{B}}{\mathcal A_{M}}  \nonumber  \\
&&~~~~~~~~~~\times \frac{f^{(n)}_{d}(x^{d}_{dsc}p_T) f^{(n)}_{s}(x^{s}_{dsc}p_T) f^{(n)}_{c}(x^{c}_{dsc}p_T)}
                      {f^{(n)}_{\bar d}(x^{d}_{dc}p_T) f^{(n)}_{c}(x^{c}_{dc}p_T)},           \label{eq:RXic0D0fin}   \\                  
%-------------------------------------------------------------------------------------------------------------------------------
&& \frac{\Xi_c^+}{D_s^+} =  \frac{3}{2+\lambda_s} \frac{\mathcal A_{B}}{\mathcal A_{M}}  \nonumber  \\
&&~~~~~~~~~~\times
                \frac{f^{(n)}_{d}(x^{d}_{dsc}p_T) f^{(n)}_{s}(x^{s}_{dsc}p_T) f^{(n)}_{c}(x^{c}_{dsc}p_T)}
                    {f^{(n)}_{\bar s}(x^{s}_{sc}p_T) f^{(n)}_{c}(x^{c}_{sc}p_T)}, ~~~~~~~~~        \label{eq:RXicpDspfin}   \\ 
%-------------------------------------------------------------------------------------------------------------------------------
&& \frac{\Omega_c^0}{D^0}  =  \frac{1.067\lambda_s^2}{2+\lambda_s} \frac{\mathcal A_{B}}{\mathcal A_{M}}
            \frac{ [f^{(n)}_{s}(x^{s}_{ssc}p_T)]^2 f^{(n)}_{c}(x^{c}_{ssc}p_T)}
                 {f^{(n)}_{\bar d}(x^{d}_{dc}p_T) f^{(n)}_{c}(x^{c}_{dc}p_T)},             \label{eq:ROmec0D0fin}   \\                   
%-------------------------------------------------------------------------------------------------------------------------------
&& \frac{\Omega_c^0}{D_s^+}  =   \frac{1.5\lambda_s}{2+\lambda_s} \frac{\mathcal A_{B}}{\mathcal A_{M}}
            \frac{ [f^{(n)}_{s}(x^{s}_{ssc}p_T)]^2 f^{(n)}_{c}(x^{c}_{ssc}p_T)}
                 {f^{(n)}_{\bar s}(x^{s}_{sc}p_T) f^{(n)}_{c}(x^{c}_{sc}p_T)}.  \label{eq:ROmec0Dspfin}                  
\end{eqnarray}  }%
We see again that, besides strangeness suppression factor $\lambda_s$ 
and ${\mathcal A_{B}}/{\mathcal A_{M}}$ (determined by the charmed baryon-to-meson ratio $R^{(c)}_{B/M}$), 
these ratios depend also on the $p_T$ spectra of quarks. 
Similarly, in the rough estimation with $x^{q_1}_{q_1q_1c}\sim x^{q_1}_{q_1q_2c}\sim x^{q_1}_{q_1c}$, 
we can rewrite them as, 
{\setlength\arraycolsep{0.2pt}
\begin{eqnarray}
&& \frac{\Lambda_c^+}{D^0} 
                      \approx \frac{4.267}{2+\lambda_s} \frac{\mathcal A_{B}}{\mathcal A_{M}}  f^{(n)}_{d}(x^{d}_{ddc}p_T)
                       \frac{ f^{(n)}_{c}(x^{c}_{ddc}p_T)} {f^{(n)}_{c}(x^{c}_{dc}p_T)},      \label{eq:RLcpD0app}   \\
%-------------------------------------------------------------------------------------------------------------------------------                     
&& \frac{\Sigma_c^0}{D^0}  
                       \approx  \frac{0.711}{2+\lambda_s} \frac{\mathcal A_{B}}{\mathcal A_{M}}  f^{(n)}_{d}(x^{d}_{ddc}p_T)
                      \frac{f^{(n)}_{c}(x^{c}_{ddc}p_T)}{f^{(n)}_{c}(x^{c}_{dc}p_T)},     \label{eq:RSc0D0app}     \\  
%-------------------------------------------------------------------------------------------------------------------------------                      
&& \frac{\Sigma_c^{++}}{D^+} 
                       \approx \frac{1.684}{2+\lambda_s} \frac{\mathcal A_{B}}{\mathcal A_{M}} f^{(n)}_{d}(x^{d}_{ddc}p_T)
                        \frac{f^{(n)}_{c}(x^{c}_{ddc}p_T)}{f^{(n)}_{c}(x^{c}_{dc}p_T)},    \label{eq:RScppDpapp}   \\                                         
%-------------------------------------------------------------------------------------------------------------------------------
&& \frac{\Xi_c^0}{D^0} 
                     \approx \frac{2.133\lambda_s}{2+\lambda_s} \frac{\mathcal A_{B}}{\mathcal A_{M}}  f^{(n)}_{s}(x^{s}_{dsc}p_T)
                     \frac{f^{(n)}_{c}(x^{c}_{dsc}p_T)}{f^{(n)}_{c}(x^{c}_{dc}p_T)},   \label{eq:RXic0D0app}  \\                  
%-------------------------------------------------------------------------------------------------------------------------------
&& \frac{\Xi_c^+}{D_s^+}   
                    \approx  \frac{3}{2+\lambda_s} \frac{\mathcal A_{B}}{\mathcal A_{M}}  f^{(n)}_{d}(x^{d}_{dsc}p_T)
                    \frac{f^{(n)}_{c}(x^{c}_{dsc}p_T)} {f^{(n)}_{c}(x^{c}_{sc}p_T)},         \label{eq:RXicpDspapp}   \\ 
%-------------------------------------------------------------------------------------------------------------------------------
&& \frac{\Omega_c^0}{D^0}  
                     =  \frac{1.067\lambda_s^2}{2+\lambda_s} \frac{\mathcal A_{B}}{\mathcal A_{M}} f^{(n)}_{s}(x^{s}_{ssc}p_T) \nonumber \\
&&~~~~~~~~~~~~~~~~~~~~~~~~~~ \times                     
            \frac{ f^{(n)}_{s}(x^{s}_{ssc}p_T) }  {f^{(n)}_{\bar d}(x^{d}_{dc}p_T)}
             \frac{f^{(n)}_{c}(x^{c}_{ssc}p_T)}{f^{(n)}_{c}(x^{c}_{dc}p_T)},             \label{eq:ROmec0D0app}   \\                   
%-------------------------------------------------------------------------------------------------------------------------------
&& \frac{\Omega_c^0}{D_s^+} 
                       \approx \frac{1.5\lambda_s}{2+\lambda_s} \frac{\mathcal A_{B}}{\mathcal A_{M}}   f^{(n)}_{s}(x^{s}_{ssc}p_T)
                        \frac{f^{(n)}_{c}(x^{c}_{ssc}p_T)}{f^{(n)}_{c}(x^{c}_{sc}p_T)}.  \label{eq:ROmec0Dspapp}                  
\end{eqnarray}  }%
Here we see clearly that these ratios, besides $\lambda_s$ and ${\mathcal A_{B}}/{\mathcal A_{M}}$, should be sensitive 
to the $p_T$ spectrum of $d$ or $s$-quarks in the relatively small $p_T$ region. 
Since the $p_T$ distribution $f^{(n)}_{q_i}(p_T)$ of $d$ or $s$-quarks in such $p_T$ regions 
typically exhibits ``rise-peak-fall'' behaviors, 
we expect that these charmed baryon-to-meson ratios should have similar rise-peak-fall behaviors.
We also expect that these ratios should have much stronger $p_T$-dependences than those of 
strange to non-strange ratios given by Eqs.~(\ref{eq:RDsD0fin}-\ref{eq:ROmecLamcfin}) since the latter 
depends only on ratios of $p_T$-spectra of quarks but the former depends the spectrum itself.
We would like to emphasize that Eqs.~(\ref{eq:RLcpD0fin}-\ref{eq:ROmec0Dspfin}) 
are characteristic results in the equal-velocity quark combination (EVC).
They can be directly used to test the combination hadronization mechanism of charm quarks and the validity of the EVC. 
They may also provide special ways to probe the properties of the QGP due to their close relationships 
to low $p_T$ spectra of $d$ and $s$ quarks. 

\section{applications in $\textrm{Au-Au}$ collisions at RHIC} \label{application}

In this section, we apply the deduced results 
in Sec.~\ref{model} to calculate the charmed hadron production at midrapidity in Au-Au collisions at $\sqrt{s_{NN}}=200$ GeV. 
We first present the normalized $p_T$-spectra of quarks and other related parameters. 
We then give numerical results of hadron yield ratios.
We finally present predictions for $p_T$ spectra and $p_T$-integrated yield densities of different charmed hadrons.

\subsection{The normalized $p_T$ spectra of quarks} 

In the midrapidity region of Au-Au collisions at high energies, 
we can neglect net quark contributions and take isospin symmetric quark distributions. 
In this case, we only need three parameters $\lambda_s$, $R^{(c)}_{B/M}$, $dN_c/dy$ and normalized
$p_T$ spectra of $d$-, $s$- and $c$-quarks as inputs. 
In QGP in heavy-ion collisions, $\lambda_s$ takes values in  
the range $0.4\sim0.6$~\cite{RQWang2012PRC}. 
%while in $pp$ and $p$-Pb reactions it is smaller, about 0.3~\cite{equal-velocity}.
Here, we use yield ratio of anti-baryons, such as $\bar \Lambda$ to $\bar p$~\cite{RQWang2012PRC,pbar2004PRC,LbarXibar2007PRL} 
to fix the value of $\lambda_s$, and the results in different centralities are shown in Table~\ref{tab_inputs}. 
The $R^{(c)}_{B/M}$ and $dN_c/dy$ will be given whenever they are needed. 

\begin{figure}[htbp]
\centering
 \includegraphics[width=1.\linewidth]{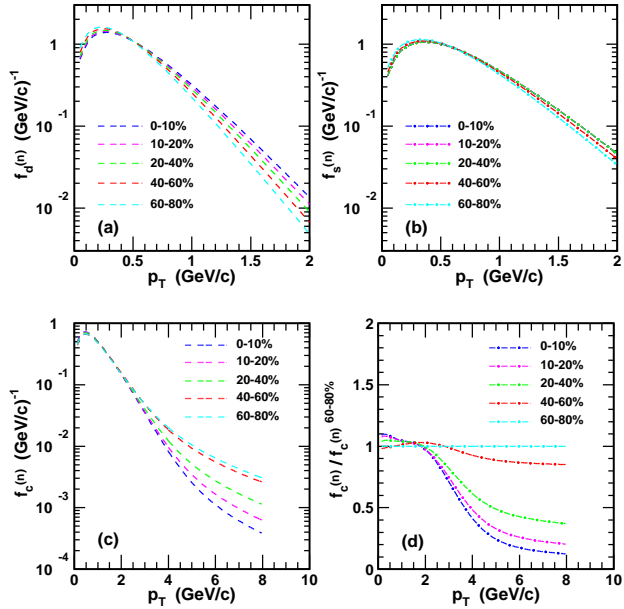}\\
 \caption{(Color online) Normalized $p_T$ spectra of (a) down, (b) strange, (c) charm quarks 
 in different centralities in Au-Au collisions at $\sqrt{s_{NN}}=200$ GeV. 
 In panel (d), we have ratios of $p_T$ distributions of charm quarks in different centralities to that in 60-80\% centrality.}
 \label{fig:quarkpt}
\end{figure}

For $f^{(n)}_{d}(p_T)$ and $f^{(n)}_{s}(p_T)$, we take the modified-thermal pattern, 
\begin{equation}
f^{(n)}_{q_i}(p_T) \propto p_T^{\alpha_{q_i}}\exp(-\sqrt{p_T^2+m_{q_i}^2}/T_{q_i}), 
\end{equation}
and extract the parameters $T_{q_i}$ and $\alpha_{q_i}$ from data~\cite{LbarXibar2007PRL,phi2007PRL} 
on the $p_T$ spectra of $\Xi^-$ baryons and $\phi$ mesons under EVC. 
The obtained results are given in Table~\ref{tab_inputs}.
We also plot $f^{(n)}_{d}(p_T)$ and $f^{(n)}_{s}(p_T)$ in Fig.~\ref{fig:quarkpt} (a) and (b).

\begin{table}[htbp]
\renewcommand{\arraystretch}{1.5}
 \centering
 \caption{$\lambda_s$ and parameters for quark distributions in different centralities in Au-Au collisions at $\sqrt{s_{NN}}= 200$ GeV.}
  \begin{tabular}{p{49pt}p{30pt}p{32pt}p{32pt}p{32pt}p{30pt}} \hline\hline
    Centrality      & 0-10\%  & 10-20\%  & 20-40\%  & 40-60\%  & 60-80\%   \\   \hline
  $\lambda_s$       &0.49     &0.46      &0.45      &0.45      &0.44      \\ \hline
  $T_d$ (GeV)       &0.27     &0.26      &0.25      &0.24      &0.23        \\
  $T_s$ (GeV)       &0.34     &0.34      &0.34      &0.33      &0.32        \\
  $\alpha_d$        &0.65     &0.65      &0.65      &0.62      &0.58        \\
  $\alpha_s$        &0.65     &0.65      &0.65      &0.62      &0.58        \\    \hline  
 $\alpha_c$ (GeV$^{-0.5}$) &11 &10        &9         &7         &6        \\    
   $T_c$ (GeV)       &0.45     &0.45      &0.45      &0.45      &0.44        \\    
  $\beta_c$         &3.65     &3.45      &3.20      &2.85      &2.80        \\  
   \hline \hline
 \end{tabular}     \label{tab_inputs}
\end{table}

For $c$-quark, we adopt the hybrid pattern~\cite{charmBTA2017PRC},  i.e.,  
\begin{eqnarray}
&&f^{(n)}_{c}(p_T)\propto\alpha_c p_T\exp(-{\sqrt{p_T^2+m_{c}^2}}/{T_c}) \nonumber\\
&&~~~~~~~~~+ \sqrt{p_T}[1.0+({\sqrt{p_T^2+m_{c}^2}-m_c})/{0.6}]^{-\beta_c}, ~~~
\end{eqnarray}
based on the results after the propagation of charm quarks in the QGP medium in a Boltzmann transport approach~\cite{charmBTA2017PRC}.
The parameters $\alpha_c$, $T_c$ and $\beta_c$ are fixed using the data on the $p_T$ distribution of $D^0$~\cite{D0PRC2019} 
and are given in Table~\ref{tab_inputs}. 
We also plot in Fig.~\ref{fig:quarkpt} (c) $f^{(n)}_{c}(p_T)$ in different centralities in Au-Au collisions at $\sqrt{s_{NN}}=200$ GeV.
We see that there is a stronger suppression in more central collisions, especially in the region 4 GeV $<p_T<$ 8 GeV.
Shown in Fig.~\ref{fig:quarkpt}(d) are the ratios of these distributions in different centralities to that in 60-80\% centrality.  
We see very similar behaviors as that of the nuclear modification factor $R_{CP}$ of $D^0$ mesons measured in~\cite{D0PRC2019}.
In low $p_T<$ 2 GeV, $p_T$ distributions of charm quarks are almost the same for different centralities.

Comparing the results in Fig.~\ref{fig:quarkpt}(c) to those given by Fig.~\ref{fig:quarkpt}(a) and \ref{fig:quarkpt}(b), 
we see that the $p_T$-dependence of $f^{(n)}_{c}(p_T)$ is much stronger than that of $f^{(n)}_{d}(p_T)$ or $f^{(n)}_{s}(p_T)$.
We therefore expect that $f^{(n)}_{c}(p_T)$ should have large influences both on $p_T$-distributions of charmed hadrons  
and on the ratios given in the last section.  

\subsection{Ratios of strange to non-strange hadrons}

\begin{figure}[htbp]
\centering
 \includegraphics[width=1.\linewidth]{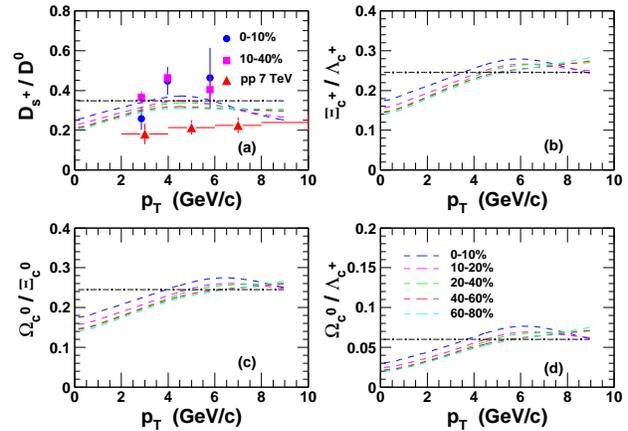}\\
 \caption{(Color online) Ratios of strange to non-strange charmed hadrons as functions of $p_T$ in Au-Au collisions at $\sqrt{s_{NN}}=200$ GeV.
 The filled cycles and squares are data that are taken from~\cite{DsD0NPA2017}. 
 The filled triangles are $pp$ reaction data obtained at LHC~\cite{DsD0pp7TeV}. 
 The dashed-dotted line in each panel represents the constant $0.711\lambda_s$, $0.5\lambda_s$, $0.5\lambda_s$ 
 or $0.25\lambda_s^2$ at $\lambda_s$=0.49, respectively.}
 \label{Ratio-strange}
\end{figure} 

With Eqs.~(\ref{eq:RDsD0fin}-\ref{eq:ROmecLamcfin}), we calculate yield ratios of strange to non-strange charmed hadrons. 
The results are shown in Fig.~\ref{Ratio-strange}.
We see that the results are basically consistent with the data~\cite{DsD0NPA2017}. 
We also see that the enhancement of $D_s^+/D^0$ in Au-Au collisions compared to those in $pp$ reactions at LHC~\cite{DsD0pp7TeV} 
comes mainly from the strangeness suppression factor $\lambda_s$ of the partonic matter in heavy ion collisions.

We also present the results for $\Xi_c^+/\Lambda_c^+$, 
$\Omega_c^0/\Xi_c^0$ and $\Omega_c^0/\Lambda_c^+$ in Fig.~\ref{Ratio-strange}.
To see effects of quark spectra in Eqs.~(\ref{eq:RDsD0fin}-\ref{eq:ROmecLamcfin}) on these ratios, 
we plot also the constant factors 
in the corresponding panel in Fig.~\ref{Ratio-strange} respectively.
We see that quark distributions slightly enhance these hadron yield ratios 
at intermediate $p_T$ region, but just the opposite at low $p_T$ area.

\subsection{Baryon-to-meson ratios}

\begin{figure}[htbp]
\centering
 \includegraphics[width=1.\linewidth]{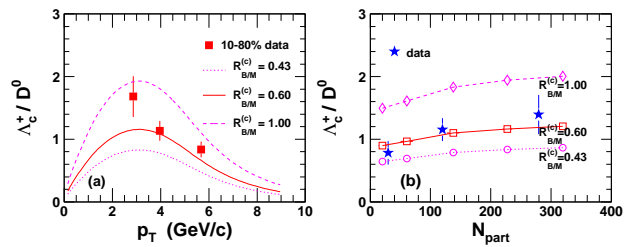}\\
 \caption{(Color online) (a) $\Lambda_c^+/D^0$ as a function of $p_T$ 
 and (b) the $p_T$-integrated $\Lambda_c^+/D^0$ as a function of $N_{part}$ in Au-Au collisions at $\sqrt{s_{NN}}=200$ GeV.
 The data are taken from~\cite{LcD0NPA2019}. 
 In (b), the open circles, squares and rhombuses connected with the dotted, solid and dashed lines to guide the eye 
 are the calculated results with different values of $R^{(c)}_{B/M}$.} 
 \label{Ratio-LcD0pt0p431p0}
\end{figure}

To calculate baryon-to-meson ratios in the charm sector,
we need the parameter $R^{(c)}_{B/M}$ to determine ${\mathcal A_{B}}/{\mathcal A_{M}}$ in Eqs.~(\ref{eq:RLcpD0fin}-\ref{eq:ROmec0Dspfin}).
It was fixed as $R^{(c)}_{B/M}=0.43$ in $pp$ and $p$-Pb reactions at LHC energies~\cite{HHLi2018PRC,JSong2018EPJC}.
In heavy ion collisions its value may be larger due to the baryon-beneficial environment.
To study the effect of $R^{(c)}_{B/M}$, we present results of calculations on $\Lambda_c^+/D^0$ in Fig.~\ref{Ratio-LcD0pt0p431p0} 
with $R^{(c)}_{B/M}=0.43$, 0.60 and 1.00, respectively.

From Fig.~\ref{Ratio-LcD0pt0p431p0}, we see, just as expected,  
that $\Lambda_c^+/D^0$ shows a quite significant $p_T$-dependence 
and the qualitative feature is consistent with the data available~\cite{LcD0NPA2019}. 
We also see that $R^{(c)}_{B/M}=0.60$ seems to be a suitable result for heavy ion collisions at top RHIC energy.
In the following calculations of this paper, we show only results obtained with this value. 

\begin{figure}[htbp]
\centering
 \includegraphics[width=1.\linewidth]{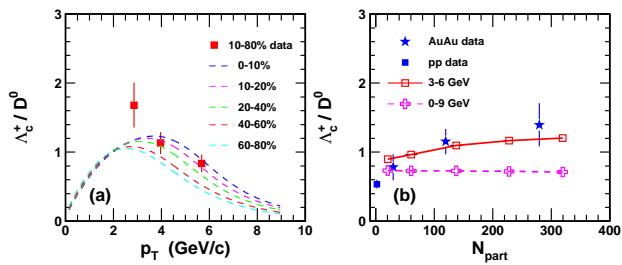}\\
 \caption{(Color online) (a) $\Lambda_c^+/D^0$ as a function of $p_T$ and 
 (b) the $p_T$-integrated $\Lambda_c^+/D^0$ as a function of $N_{part}$ in Au-Au collisions at $\sqrt{s_{NN}}=200$ GeV. 
 The data are taken from~\cite{LcD0NPA2019,pppPb2018JHEP}. 
In (b) the Au-Au data are for $p_T$ in the range from 3 to 6 GeV and that for $pp$ is for $p_T$ in the range from 3 to 4 GeV.}
 \label{Ratio-LcD0ptNpart}
\end{figure}

To further study influences of quark distributions on baryon productions, 
we calculate $p_T$ dependence of $\Lambda_c^+/D^0$ in different centralities in Au-Au collisions at $\sqrt{s_{NN}}=200$ GeV.
The results are given in Fig.~\ref{Ratio-LcD0ptNpart}. 
We see that they all exhibit similar rise-peak-fall behaviors.
From central to peripheral collisions, peak values decrease from about 1.3 to 1.0 and the locations shift to lower $p_T$.
This is due to the centrality dependence of $p_T$-distributions of quarks shown in Fig.~\ref{fig:quarkpt}, especially charm quarks.
Other models such as the Catania model~\cite{RLcD2018EPJC} 
that includes coalescence and fragmentation gives also the rise-peak-fall behavior of $\Lambda_c^+/D^0$ as the function of $p_T$. 
However, the Catania model~\cite{RLcD2018EPJC} predicts much flatter rising behavior at low $p_T$ region 
and almost no shift of peak location from RHIC to LHC energies.
Experimental measurements can distinguish these different models 
and provide important insights into charm quark hadronization in high energy collisions.

Data of the $p_T$-integrated $\Lambda_c^+/D^0$ from central to peripheral Au-Au collisions at $\sqrt{s_{NN}}=200$ GeV 
show a decreasing trend~\cite{LcD0NPA2019} that is different from in light sectors~\cite{FLShao2017PRC} 
where baryon-to-meson ratios show little centrality dependence. 
We see that the calculated results in the same $p_T$ range (open squares in the figure) 
exhibit indeed such a trend consistent with the data~\cite{LcD0NPA2019}.
However, if we take $p_T$ integrated from 0 GeV to 9 GeV (shown by the open crosses in the figure), 
this trend disappears and the result is essentially independent of the centrality and lower than those integrated from 3 GeV to 6 GeV.
Such properties come from the centrality dependence of $c$-quark distributions given by Fig.~\ref{fig:quarkpt} 
where we see that a strong dependence for larger $p_T$ but negligible in the small $p_T$ region. 
The results for $p_T$ integrated from 0 to 9 GeV are dominated by the small $p_T$ contributions. 

\begin{figure}[htbp]
\centering
 \includegraphics[width=1.\linewidth]{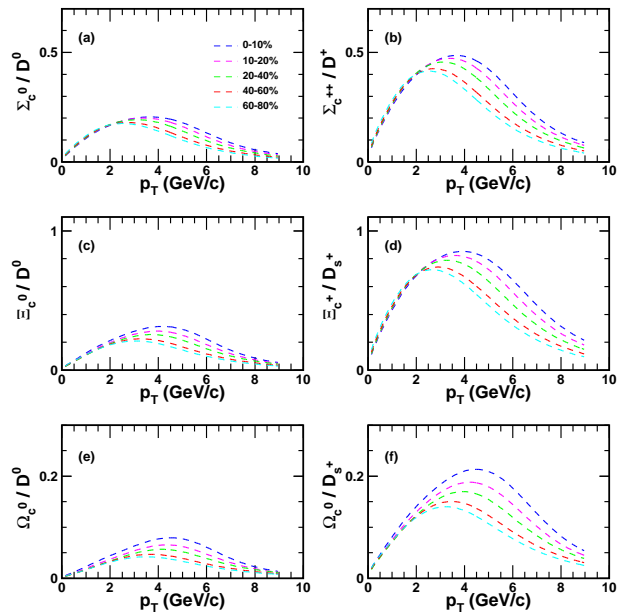}\\
 \caption{(Color online) Different baryon-to-meson ratios as functions 
 of $p_T$ in different centralities in Au-Au collisions at $\sqrt{s_{NN}}=200$ GeV. }
 \label{Ratio-BcDpt}
\end{figure}

\begin{figure}[htbp]
\centering
 \includegraphics[width=1.\linewidth]{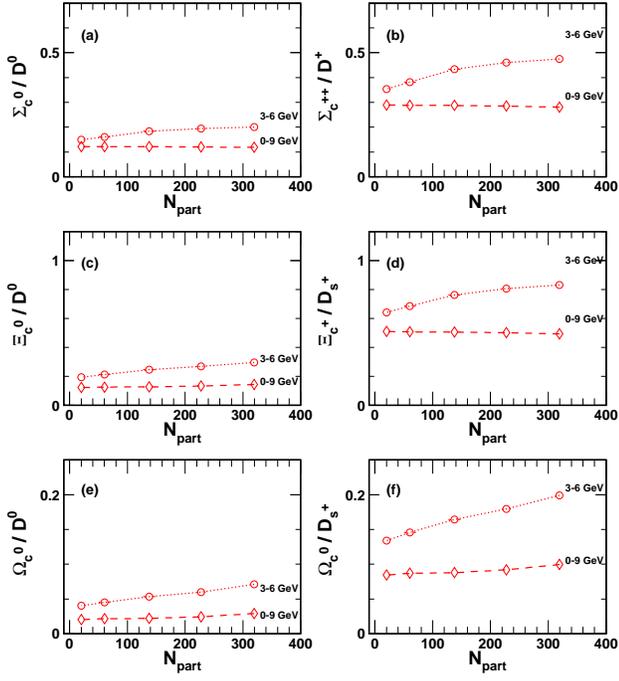}\\
 \caption{(Color online) Different $p_T$-integrated baryon-to-meson ratios 
 as the function of $N_{part}$ in Au-Au collisions at $\sqrt{s_{NN}}=200$ GeV.
 Open circles are for $p_T$ integrated from $3$ to $6$ GeV and open rhombuses are from $0$ to $9$ GeV.}
 \label{Ratio-BcDNpart}
\end{figure}

Encouraged by the agreements with data available, 
we make predictions for other similar baryon-to-meson ratios. 
The results are given in Figs.~\ref{Ratio-BcDpt} and~\ref{Ratio-BcDNpart}. 

In Fig.~\ref{Ratio-BcDpt}, we see that all the ratios exhibit similar rise-peak-fall behaviors as functions of $p_T$, 
and the peak locations change from central to peripheral collisions similar to $\Lambda_c^+/D^0$.
In Fig.~\ref{Ratio-BcDNpart}, 
we see similar trend for all these $p_T$-integrated ratios, i.e., they all show increasing tendencies for results integrated 
in the $p_T$ region from 3 to 6 GeV but almost flat if integrated from 0 to 9GeV.  

At the end of this part, we would also like to emphasize that in our calculations, not merely $\Lambda_c^+$, 
but all charmed baryons are enhanced according to the overall ratio $R^{(c)}_{B/M}$ that was taken as $0.60$ in the calculations.
This is different from other coalescence models where diquarks were introduced to intensely 
enhance the production of $\Lambda_c^+$, but to less enhance 
or even not enhance other charmed baryons~\cite{Lamc-diquark}.
Future measurements of different charmed baryons should be very helpful in distinguishing different models and understanding the enhancment mechanism of charmed baryon production.   

\subsection{The $p_T$ spectra of charmed hadrons}

Having the $p_T$ distributions of quarks, we not only calculate the ratios presented above 
but also the $p_T$-spectra of charmed hadrons obtained under EVC.  We present the results in the following. 
To obtain not only the shape but also the magnitudes, 
we need the rapidity density of charm quarks $dN_c/dy$ ( at $y=0$) as an input.
For this purpose, we estimate it by extrapolating $pp$ reaction data on differential cross section 
${d\sigma^{pp}_{c}}/{dy}$ and take $dN_{c}/dy=\langle T_{AA}\rangle{d\sigma^{pp}_{c}}/{dy}$, 
where $\langle T_{AA}\rangle$ is the average nuclear overlap function and can be calculated by the Glauber model~\cite{Glauber1970NPB,TAA2018PRC}. 
We use ${d\sigma^{pp}_{c}}/{dy}=130\pm30\pm26$ $\mu b$ recently measured at midrapidity 
in $pp$ at $\sqrt{s}= 200$ GeV~\cite{LcD0NPA2019}, 
and obtain $dN_c/dy=2.945\pm0.680\pm0.589$ in the most central 0-10\% collisions in Au-Au collisions at $\sqrt{s_{NN}}= 200$ GeV.
Considering the data of $D^0$~\cite{D0PRC2019}, 
we take $dN_c/dy=2.45$ for the centrality 0-10\%. 
For other centralities 10-20\%, 20-40\%, 40-60\%, 60-80\%, we have $dN_c/dy=$ 1.54, 0.76, 0.24, 0.055, respectively.

\begin{figure}[htbp]
\centering
 \includegraphics[width=1.\linewidth]{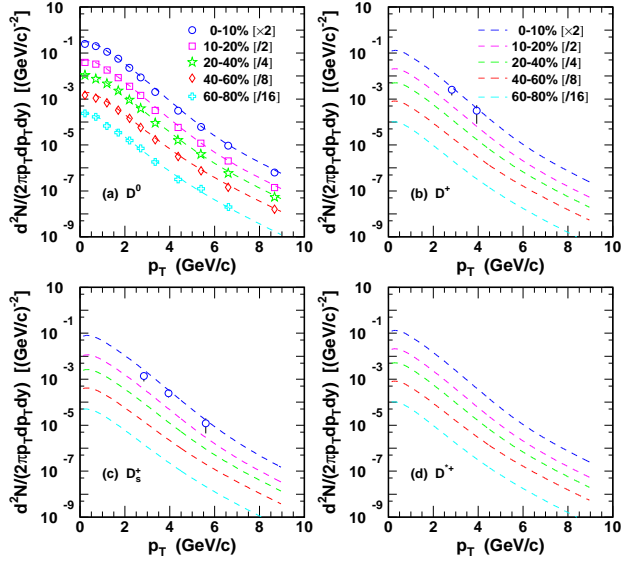}\\
 \caption{(Color online) $p_T$ spectra of open charm mesons in Au-Au collisions at $\sqrt{s_{NN}}=200$ GeV in different centralities.
 The open symbols are data taken from~\cite{D0PRC2019,DsD0NPA2017}.}
 \label{fig:Dpt}
\end{figure}

\begin{figure}[htbp]
\centering
 \includegraphics[width=1.\linewidth]{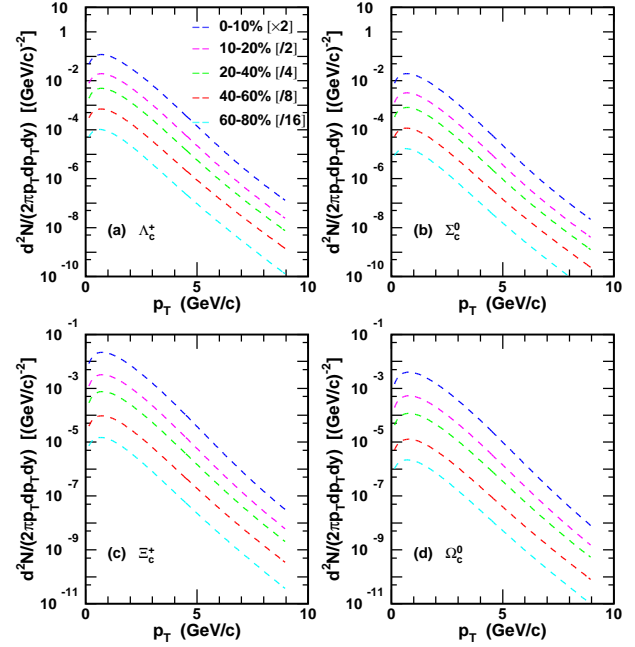}\\
 \caption{(Color online) Calculated results for $p_T$ spectra of singly charmed baryons 
 in different centralities in Au-Au collisions at $\sqrt{s_{NN}}=200$ GeV.}
 \label{fig:Bcpt}
\end{figure}

In Fig.~\ref{fig:Dpt}, we present $p_T$ spectra of different $D$ mesons in Au-Au collisions at $\sqrt{s_{NN}}=200$ GeV 
together with data available~\cite{D0PRC2019,DsD0NPA2017}. 
We see that the results agree reasonably with the data. 
In fig.~\ref{fig:Bcpt}, we show results for different charmed baryons. 
We also present $p_T$-integrated yield densities $dN/dy$ of different charmed hadrons 
at the midrapidity in different centralities in Table \ref{tab_yields}.
These results can all be used to test the mechanisms in particular EVC by future experiments.

\begin{table}[htbp]
\renewcommand{\arraystretch}{1.5}
 \centering
 \caption{Yield densities $dN/dy$ of charmed hadrons in the midrapidity in different centralities in Au-Au collisions at $\sqrt{s_{NN}}= 200$ GeV.}
  \begin{tabular}{p{30pt}p{35pt}p{35pt}p{35pt}p{35pt}p{35pt}}
    \hline\hline
    Hadron          & 0-10\%  & 10-20\%  & 20-40\%  & 40-60\%  & 60-80\%   \\
    \hline
  $D^0$             &0.893    &0.570     &0.284     &0.0898    &0.0207        \\
                    
  $D^+$             &0.377    &0.241     &0.120     &0.0379    &0.00874        \\

  $D^{*+}$          &0.381    &0.243     &0.121     &0.0383    &0.00883        \\

  $D_s^+$           &0.261    &0.151     &0.0713    &0.0223    &0.00495       \\

  $\Lambda_c^+$     &0.636    &0.412     &0.207     &0.0655    &0.0152        \\
  
  $\Sigma_c^0$      &0.106    &0.0687    &0.0344    &0.0109   &0.00253        \\

  $\Sigma_c^{++}$   &0.106    &0.0687    &0.0344    &0.0109    &0.00253       \\
  
  $\Xi_c^0$         &0.129    &0.0758    &0.0360    &0.0113    &0.00252        \\
  
  $\Xi_c^+$         &0.129    &0.0758    &0.0360    &0.0113    &0.00252       \\
    
  $\Omega_c^0$      &0.0260   &0.0139    &0.00626   &0.00194   &0.000419        \\               
    \hline \hline
 \end{tabular}     \label{tab_yields}
\end{table}

\section{summary}

Though not much data available yet, charm hadron production seems to provide an important test 
of different hadronization mechanisms in heavy ion collisions. 
In this paper, we have derived the $p_T$-dependence 
of open charm mesons and singly charmed baryons in the quark combination model under the EVC 
in ultra-relativistic heavy ion collisions.
We present in particular analytic expressions of two groups of hadron yield ratios, 
the strange to non-strange charmed hadron ratios and baryon-to-meson ratios 
in terms of normalized $p_T$ spectra of quarks. 
We present normalized $p_T$ spectra of quarks and numerical results for these hadron yield ratios using 
these quark $p_T$ spectra. 
We found that the magnitude of the strange to non-strange charmed hadron ratios are mainly determined 
by the strangeness suppression factor and have weak $p_T$ dependences. 
In contrast, there is an obvious $p_T$ dependence for baryon-to-meson ratios determined by the quark $p_T$ spectra. 
The different baryon-to-meson ratios have similar $p_T$ and centrality dependences sensitive to $p_T$ distribution of $c$-quark.
We have compared the results obtained with the data available and present predictions for future experiments.  
Further studies along this line can provide more sensitive tests to charm quark hadronization mechanisms 
and insight on properties of the QGP in heavy ion collisions.

\section*{Acknowledgements}

We thank Zhang-Bu Xu for helpful discussions.
This work was supported in part by the National Natural Science Foundation of China under grant 11505104, 11575100, 11675092 and 11975011 and by the Natural Science Foundation of Shandong Province, China under grant ZR2019YQ06.

\end{document}